\documentclass[lettersize,journal]{IEEEtran}
\usepackage{cite}
\usepackage{amsmath, amssymb, amsfonts}
\usepackage{graphicx}
\usepackage{textcomp}
\usepackage{xcolor}
\usepackage{bm}
\usepackage{enumerate}
\usepackage{threeparttable}
\usepackage{tabularx} 
\usepackage{float}
\usepackage{caption}
\usepackage{placeins}
\usepackage{latexsym}
\usepackage{subcaption}
\usepackage{wrapfig}
\usepackage{lscape}
\usepackage{pgfgantt}
\usepackage{times}
\usepackage{stfloats}
\usepackage{multirow}
\usepackage{xspace}
\usepackage{calc}
\usepackage{fancyhdr}
\usepackage{varioref}
\usepackage{tikz}
\usepackage{booktabs}
\usepackage{theorem}
\usepackage{algorithm}
\usepackage{algcompatible}
\usepackage{longtable}
\usepackage{verbatim}
\usepackage{algpseudocode}
\usepackage{url}
\usepackage{colortbl}
\usepackage{enumitem}
\usepackage{amsmath}
\usepackage{bm}
\usepackage{threeparttable}
\usepackage{makecell}
\usepackage{soul} 
\usepackage{tabularx} 
\usepackage{hhline}
\usepackage{boldline} 
\usepackage{tcolorbox}
\usepackage{pifont}

\newcommand{\inlineitem}[1][]{%
  \ifnum\value{enumi}>0
    \ifnum\value{enumii}=0
      \refstepcounter{enumi}%
    \else
      \refstepcounter{enumii}%
    \fi
    \textbf{\theenumi.} #1\hspace{\labelsep}%
  \fi
}

\newcommand{\cmark}{\ding{52}}%
\newcommand{\xmark}{\ding{54}}%

\fancypagestyle{firstpage}{
    \fancyhf{}
    
    \fancyhead[C]{\small This work has been submitted to the IEEE for possible publication. Copyright may be transferred without notice, after which this version may no longer be accessible.}
}

\begin{document}
\title{Privacy-Preserving and Trustworthy \\ Deep Learning for Medical Imaging \thanks{This work has been submitted to the IEEE for possible publication. Copyright may be transferred without notice, after which this version may no longer be accessible.}} 

\author{
  \IEEEauthorblockN{Kiarash Sedghighadikolaei, Attila A Yavuz}\\
  \IEEEauthorblockA{\textit{\small{Computer Science \& Engineering, University of South Florida, Tampa, USA}} \\
  \{kiarashs,attilaayavuz\}@usf.edu}
}

\maketitle

\begin{abstract}
The shift towards efficient and automated data analysis through Machine Learning (ML) has notably impacted healthcare systems, particularly Radiomics. Radiomics leverages ML to analyze medical images accurately and efficiently for precision medicine. Current methods rely on Deep Learning (DL) to improve performance and accuracy (Deep Radiomics). Given the sensitivity of medical images, ensuring privacy throughout the Deep Radiomics pipeline—from data generation and collection to model training and inference—is essential, especially when outsourced. Thus, Privacy-Enhancing  Technologies (PETs) are crucial tools for Deep Radiomics. Previous studies and systematization efforts have either broadly overviewed PETs and their applications or mainly focused on subsets of PETs for ML algorithms. In Deep Radiomics, where efficiency, accuracy, and privacy are crucial, many PETs, while theoretically applicable, may not be practical without specialized optimizations or hybrid designs. Additionally, not all DL models are suitable for Radiomics. Consequently, there is a need for specialized studies that investigate and systematize the effective and practical integration of PETs into the Deep Radiomics pipeline. This work addresses this research gap by (1) classifying existing PETs, presenting practical hybrid PETS constructions, and a taxonomy illustrating their potential integration with the Deep Radiomics pipeline, with comparative analyses detailing assumptions, architectural suitability, and security, (2) Offering technical insights, describing potential challenges and means of combining PETs into the Deep Radiomics pipeline, including integration strategies, subtilities, and potential challenges, (3) Proposing potential research directions, identifying challenges, and suggesting solutions to enhance the PETs in Deep Radiomics.

\end{abstract}

\begin{IEEEkeywords}
Deep Learning; Healthcare; Radiomics; Privacy-enhancing technologies; Trustworthy Artificial Intelligence.
\end{IEEEkeywords}

\section{Introduction}\label{sec:introduction}
\IEEEPARstart{A}{s} traditional technologies are replaced by digital alternatives, the digital data generation has significantly increased \cite{rani2023integration}. This is particularly evident in the medical sector with the digitalization of clinical information into Electronic Health Records (EHRs) \cite{huang2020fusion}. While there are various medicine study fields frequently generating extensive data like genomics \cite{jiang2022big} and proteomics \cite{cui2022high}, the National Institutes of Health (NIH) has identified Radiomics \cite{jha2023emerging} as one of the promising technologies for quantitatively analyzing high-dimensional medical imaging data, applicable to various modalities like Magnetic Resonance Imaging (MRI), Computed Tomography (CT), and nuclear medicine imaging, etc.

The increasing data volume and sophistication of image processing and querying for Radiomics pose storage and maintenance challenges. While cloud storage services offer cost-effective solutions, storing and managing sensitive EHR data on remote servers brings significant privacy and security concerns \cite{ehrsurveysecurity} such as cloud data breaches \cite{nihcouldbreach}, regulations and law \cite{HIPAA}, and malware infections \cite{cbsnewsAscensionHealthcare,wiredMedicalTargetedRansomware,nbcnewsRansomwareAttack}. Encrypting data before outsourcing to the cloud storage ensures confidentiality, but conventional encryption methods (e.g., AES \cite{AES}) are insufficient for frequent search and update operations on medical records (e.g., a hospital accessing patients' records), requiring full dataset downloads for decryption and causing high communication overhead. Moreover, encryption does not provide authentication for the downloaded dataset. Hence, encrypted search mechanisms with integrity assurance are vital for Deep Radiomics to ensure data privacy while supporting efficient search and update operations. 

While the secure storage, access, and management of Radiomics data can be (in part) addressed with encrypted search mechanisms, traditional data processing methods, like hospitals examining CT images for statistical studies, are time-consuming, error-prone, and potentially inaccurate. Machine Learning (ML), particularly Deep Learning (DL), offers a promising solution for extracting structural knowledge from datasets and predicting outcomes accurately. DL has proven to be effective in Radiomics \cite{fromMLtoDeep,TraditionalMLtoCNN,gupta2023fedcare}, which can benefit {precision medicine} \cite{chen2022radiomics} to identify patient subtypes to achieve optimal outcomes and accurately predict prognosis \cite{bakas2020overall}. This is vital as, for instance, in tumor detection, patients with similar histopathological features in medical images may have varying survival probabilities \cite{hagele2020resolving}. Achieving satisfactory precision necessitates DL having access to substantial computational resources (e.g., hardware accelerators \cite{nvidiaNVIDIA6000,tpu}), often provided by cloud services like Amazon AWS \cite{amazonMachineLearning} and Microsoft Azure \cite{microsoftAzureMachine}. While platforms like Microsoft Azure at UCLA Health for EHR data integration \cite{uclahealth} and Google AI at Mayo Clinic for breast cancer risk scoring \cite{mayoclinic} are used in healthcare, outsourcing sensitive data for computation poses privacy risks, which even high security encrypted search mechanisms are insufficient due to their potential leakages. Additionally, achieving high accuracy through multi-institutional training, such as inter-hospital training \cite{multiinstdeeplearningnih}, presents architectural challenges. Therefore, employing methods that ensure privacy in various architectural settings while enabling DL integration with Radiomics in potentially untrusted environments is crucial.

{\em \underline{Objective and Scope:}} 
While various Neural Networks (NNs) are used for different EHRs (e.g., Recurrent Neural Networks (RNNs) \cite{graves2013speech} for blood pressure and EEG), to use DL in Radiomics (Deep Radiomics), this paper focuses on Convolutional Neural Networks (CNNs) \cite{krizhevsky2012imagenet} for their superior pattern recognition in medical imaging \cite{sarvamangala2022convolutional}. CNNs efficiently extract high-order hierarchical features, making them ideal for Radiomics \cite{robbins1951stochastic,DLforRadiomicsReview,DLRadiomicsBrainTumor,DLforRadiomicsCancerDiag}. In addition, training CNNs with Stochastic Gradient Descent (SGD) \cite{sgdvariants} ensures rapid convergence and computational efficiency, which benefits Deep Radiomics.

While several surveys and systemizations of knowledge exist that have addressed PETs and their technical aspects (e.g., \cite{sseli2023survey,mpcorsini2021efficient,pirvithana2023private,oramsurvey,dpyang2023local,fhemarcolla2022survey,zkpmorais2019survey,flbanabilah2022federated}), as well as the intersection of PETs and ML as Privacy-preserving Machine Learning (PPML) (e.g., \cite{npsmart,qin2023cryptographic,ng2023sok,SokPPML,ppmlacmsurvey,darzi2024pqcmeetsml,ppmlpqc}), to the best of our knowledge, this study is the first to comprehensively systematize, categorize, and analyze PETs applied to Deep Radiomics as Privacy-preserving Deep Radiomics (PPDR). We focus on the properties, application potential, hybrid combinations, and pros and cons of PETs rather than their technological details (e.g., formal definitions). 

Below, we outline the research gap and our contributions.

\vspace{-3.5mm}
\subsection{Research Gap}
Several studies have explored PPML from different directions, including ML \cite{ppmlcheng,Xu2021PrivacyPreservingML,SokPPML,ppmlacmsurvey}, cryptographic techniques \cite{qin2023cryptographic}, and post-quantum secure cryptographic methods \cite{darzi2024pqcmeetsml,ppmlpqc}. Regarding private search, recent studies have reviewed Searchable Encryption (SE) \cite{song2000practical}, assessing its performance and vulnerabilities \cite{sseli2023survey,sebosch2014survey}, Private Information Retrieval (PIR) \cite{pirchor} and its diverse adaptations and extensions \cite{pirvithana2023private}, and Oblivious Random Access Machine (ORAM) \cite{oramsurvey}. For computations under encryption, Multi-Party Computation (MPC) \cite{yao1982protocols} was investigated by \cite{mpc2020lindell,lindell2005secure,mpcorsini2021efficient} in analytics and government collaboration. Moreover, Zhou et al. \cite{mpcdlzhou2024secure} showed how MPC integrates with other PETs and works \cite{mpcdlzhang2021privacy,ng2023sok} explored MPC methodologies tailored for DL and the effectiveness of MPC-based protocols for private inference and training was investigated by \cite{mpcfeng2022concretely}. Furthermore, the role of Homomorphic Encryption (HE) \cite{chillotti2020tfhe} in cloud computing, aggregation, private database queries, and bioinformatics was explored by \cite{fhemarcolla2022survey,fhemartins2017survey,wood2020homomorphic}. Acar et al. \cite{fheacar2018survey} showed theoretical and hardware implementations of HE, while works in \cite{hepodschwadt2022survey,ppdlhe,ng2023sok} investigated HE applications in PPDL. Works in \cite{femascia2021survey,fexu2020revisiting} investigated various Functional Encryption (FE) \cite{feformalized} schemes, outlining their features, constraints, and security architectures. Panzade et al. \cite{fedlpanzade2023privacy} extensively analyzed diverse strategies for FE-based PPML, highlighting their strengths and weaknesses. Hybrid constructions of these PETs were investigated by \cite{npsmart}, where integrating with Differential Privacy (DP) \cite{dwork2006differential} was studied by \cite{ppdlsurvtanuwidjaja2020privacy}, and with Trusted Execution Environments (TEEs) \cite{intelsgx} by \cite{ppdlcabrero2021sok}. These integrations were further thoroughly reviewed and categorized by \cite{ppdlsurvboulemtafes2020review,ppdlmireshghallah2020privacy}.

The works in~\cite{zkpsun2021survey,zkpmorais2019survey} delved into verifiable execution techniques like Zero-Knowledge Proofs (ZKPs) \cite{goldwasser2019knowledge}, highlighting their relevance in Blockchain and Cryptocurrency transactions. Furthermore, Sathe et al. \cite{zkpdlsathe2023state} analyzed ZKP for inference in PPML and compared the efficiency of the prior schemes, whereas \cite{zkpdlxing2023zero} provided a comprehensive classification of ZKP methods employed in PPML. TEEs were comprehensively presented by Hosam et al. \cite{teehosam2022comprehensive}, and Valadares et al. \cite{teevaladares2021systematic} illustrated their utility in Cloud/Fog-Based IoT. The TEE's role in ML was presented by \cite{teedlmo2022machine,teeli2023survey}. Applying Federated Learning (FL) \cite{adnan2022federated} across various domains, including recommender systems, IoT, and the medical, has been examined by \cite{kaissis2020MagMedicalImage,flbanabilah2022federated,flzhang2021survey}, and collaborative DL during the training stage was classified by \cite{ppdlantwi2021privacy}. Lastly, DP has been researched for statistical perturbation, focusing on its application in statistical queries \cite{dpyang2023local} and learning tasks \cite{dpdlel2022differential}. Blanco-Justicia et al. \cite{dpdlblanco2022critical} critically evaluated previous DP-based PPML methods, highlighting risks linked to privacy parameter misuse.

Previous investigations regarding taxonomy, survey, and categorization on PETs (and ML) can be grouped into two classes: Firstly, studies that provide comprehensive introductions to PETs, discussing their applications across diverse domains. Secondly, studies exploring the intersection of PETs with ML, with some focusing on specific subsets of PETs and their integrations, while others offer broader overviews across PETs and various ML models. While all PETs theoretically can be applied to Deep Radiomics, practical limitations due to their underlying constructs may hinder their effectiveness. Therefore, effectively integrating PETs into the Deep Radiomics pipeline requires a specialized study as a roadmap to address challenges and present future directions toward practical PPDR, which is currently missing in the literature.

\vspace{-3mm}
\subsection{Our Contribution} \label{subsec:Contrib}
We aim to fill the gap in the literature with the following contributions:

\begin{figure*}[htbp]
    \centering
    \includegraphics[width=0.95\textwidth]{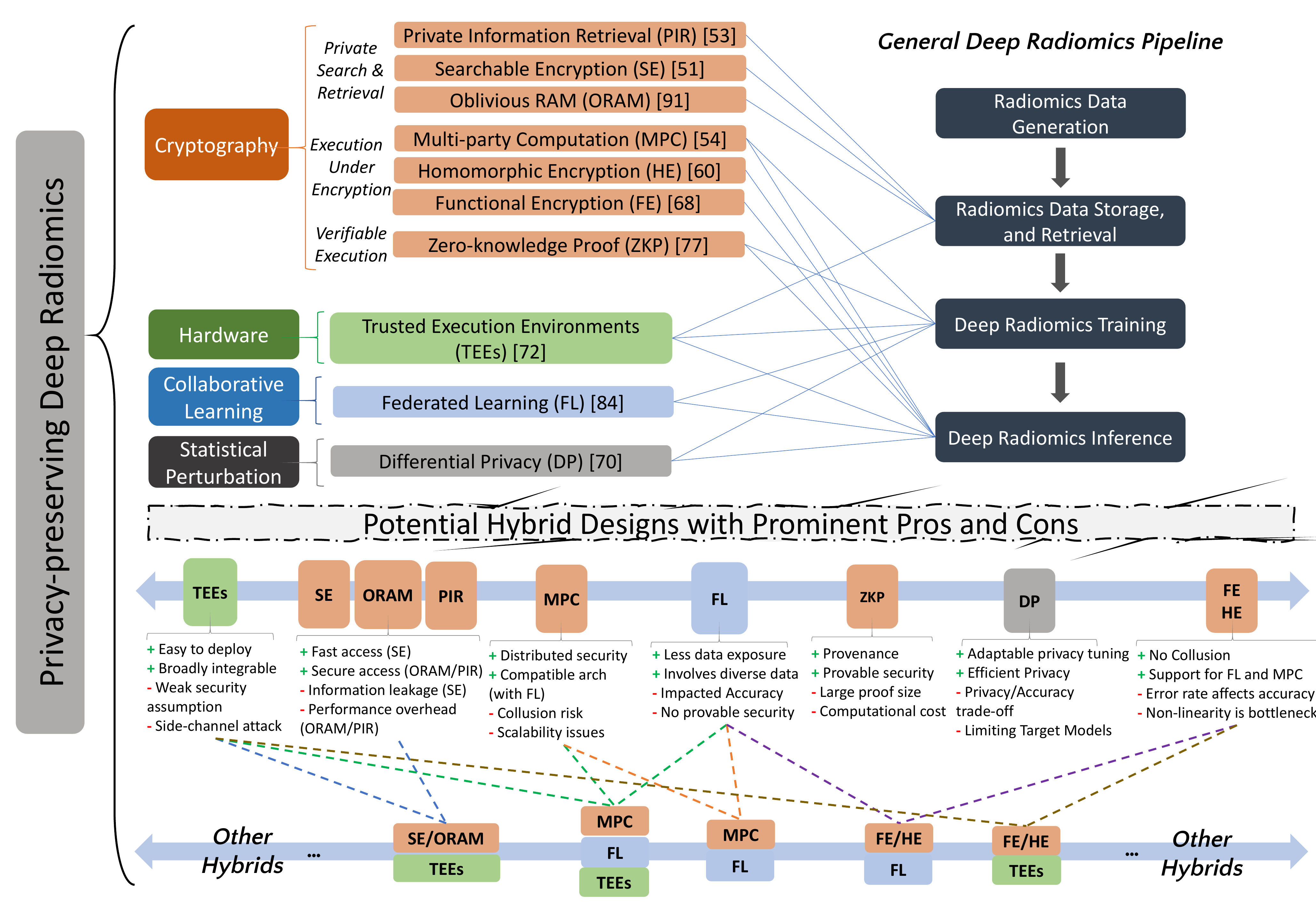} 
    \caption{Taxonomy, Pros, and Cons of Essential PETs with their Prominent Hybrid Compositions for Deep Radiomics}
    \label{fig:ppmlclassifcation}
    \vspace{-5mm}
\end{figure*}

\vspace{1mm}
\noindent $\bullet$ {\em \ul{Detailed Taxonomy and Classification of Prominent PETs for Deep Radiomics with Hybrid Compositions in Mind}}: 
We categorized various PETs applicable to Deep Radiomics into four groups based on their fundamental principles. While each PET holds theoretical applicability, practical implementation often requires integrating them with other PETs to create hybrid designs. Hence, we also present practical hybrid designs suitable for real-world deployment. For this matter, Fig. \ref{fig:ppmlclassifcation} illustrates our primary PETs classification for the Deep Radiomics pipeline, detailing their respective advantages, disadvantages, and prominent hybrid configurations. Additionally, TABLE \ref{tab:petcomp} lists these PETs and their significant hybridizations, offering detailed insights into their performance, security, architectural suitability, underlying assumptions, etc.

\begin{figure*}[!t]
    \centering
    \includegraphics[width=\textwidth]{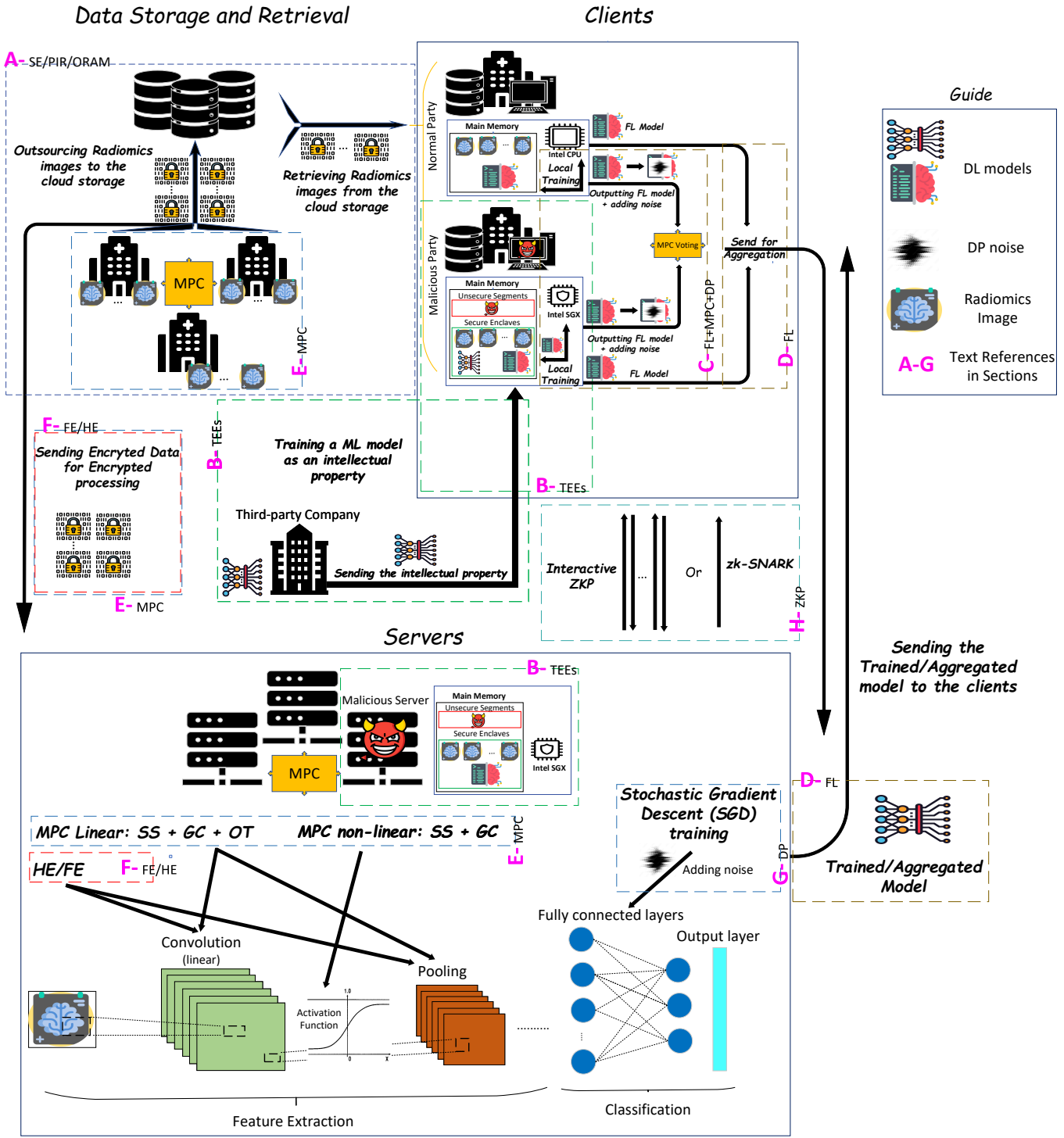}
    \caption{Integration of PETs with Deep Radiomics Pipeline.}
    \label{fig:mlpipeline}
    \vspace{-6mm}
\end{figure*}

\vspace{1mm}
\noindent $\bullet$ {\em \ul{Comprehensive yet Broadly Accessible Technical Insights on Deep Radiomics's Pipeline with PETs}}:  Given the necessity of considering the full pipeline of Deep Radiomics, our examination spans the stages of data storage/retrieval phases, model training, and model release and inference. Having such detailed classification and scrutinizing the essential prerequisites of each stage, we present a thorough yet broad technical analysis of incorporating each PET into each stage (Fig. \ref{fig:mlpipeline}). This analysis presents the performance challenges and possible mitigations against privacy attacks in Deep Radiomics.

\noindent $\bullet$ {\em \ul{Presenting a Roadmap Towards Privacy-preserving Deep Radiomics with Challenges and Future Research Directions}}:
Our study offers a potential roadmap for integrating PETs throughout the entire Deep Radiomics pipeline, focusing on practical deployment considerations. Each PET is evaluated in terms of its main advantages, integration challenges, and functional limitations. Additionally, we discuss future directions to address these challenges in each stage, including enhancements required for each PET and exploring effective hybrid design solutions.

{\em Organization}: Section II presents private search and data access. Section III explores execution under encryption, while Section IV investigates verifiable execution. The role of collaborative and distributed learning is discussed in Section V, followed by the significance of trusted hardware environments in Section VI and statistical perturbation in Section VII.


\begin{table*}[ht]

  \caption{Overview of Privacy-Preserving Technologies for Deep Radiomics $^\dag$}
  \label{tab:petcomp}
  \centering
  \begin{tabularx}{\textwidth}{>{\centering\arraybackslash}p{0.1cm}>{\arraybackslash}m{1.4cm}>{\centering\arraybackslash}m{1cm}>{\centering\arraybackslash}m{0.9cm}>{\centering\arraybackslash}m{0.8cm}>{\centering\arraybackslash}m{0.8cm}>{\centering\arraybackslash}m{1.2cm}>{\centering\arraybackslash}m{0.6cm}>{\centering\arraybackslash}m{0.6cm} >{\arraybackslash}m{3cm}>{\arraybackslash}m{5cm}}
    \toprule
    \textbf{Type \footnote{}} & \centering \textbf{PET} & \vspace{1mm}\textbf{Encrypted \newline Comp.} & \vspace{1mm} \textbf{Training/\newline Inference} & \vspace{1mm} \textbf{Training Time} & \vspace{1mm} \textbf{Inference Time} & \vspace{1mm} \textbf{Accuracy} & \vspace{1mm} \textbf{Comp. \newline Cost} & \vspace{1mm} \textbf{Comm. \newline Cost} & \centering \textbf{Security} & \hspace{3mm} \textbf{Special Utility} \\ 
    \midrule
    \multirow{7}{*}{\textbf{\vspace{-8mm} \hspace{-1mm}C}} &\centering  MPC \cite{yao1982protocols} & \cmark & \cmark/\cmark & M & M & Good \footnote{} & M & H & Semi-honest/Malicious/ \newline $(t, n)$ Malicious/ \newline Collusion \footnote{} & OT: Non-linear operations \newline SS/GC: Linear and non-linear\\ 
    
    & \centering ZKP \cite{goldwasser2019knowledge} & \xmark &  \cmark/\cmark & M-H & M-H & Avg-Good & H & H & ZKP's properties \footnote{} & Non-repudiation  \\ 

    & \centering HE \cite{feformalized} & \cmark &  \xmark \footnote{}/\cmark & H & H & Good & H & L & No collusion \footnote{} & Linear friendly\\

    & \centering FE \cite{gentry2009fully} & \cmark &  \xmark\footnote[5]{}/\cmark & H & H & Good & H & L & No collusion & Linear friendly \\
    \cmidrule(l{-16mm}r{-65mm}){3-9}
    
    & \centering SE \cite{song2000practical} & \xmark &  N/A & N/A & N/A & N/A & L & L & Information leakage \footnote{SE} & Versatile search capabilities\\
    
    & \centering ORAM \cite{goldreich1987towards} & \xmark &  N/A & N/A & N/A & N/A & M & M & Obliviousness \footnote{} & Hiding access pattern\\

    & \centering PIR \cite{pirchor} & \xmark &  N/A & N/A & N/A & N/A & L-M & M-H & Oblivious retrieval & Hiding query\\
    \midrule

    \multirow{1}{*}{\hspace{-1mm} \textbf{H}} & \centering TEE \cite{intelsgx} & \cmark &  \cmark/\cmark & L & L & Good \footnote{} & L \footnote{}& L & Side-channel susceptible \newline No provable security & Broad Integrability and \newline Attestation \\
    \midrule    
    
    \multirow{1}{*}{\hspace{0.4mm}\textbf{L}} & \centering FL \cite{li2020multi} & \xmark &  \cmark/\xmark & M \footnote{} & M \footnote{}& Good \footnote{} & M & M & No provable security & Reduced Data Exposure \\
    \midrule

    \multirow{1}{*}{\hspace{0.4mm}\textbf{S}} & \centering DP \cite{dwork2006differential} & \xmark &  \cmark/\xmark \footnote{} & M & N/A \footnote[14]{} & Avg \footnote[14]{} & L & L & Statistical security & Adaptable privacy tuning\\

    \midrule[\doublerulesep]
    \multirow{5}{*}{\parbox{2cm}{\rotatebox[origin=c]{90}{\centering \hspace{-5mm}\textbf{Hybrid}}}} & \centering MPC+TEE \cite{choi2019hybrid}& \cmark &  \cmark/\cmark  & M & M & Good & M & H & Malicious \newline Side-channel risk & Distributed malicious network\\ 
    & \centering MPC+FL \cite{kaissis2021end}& \cmark &  \cmark/\cmark & M & M & Good & H & H & Same as MPC & Collaborative FL inference\\
    & \centering TEE+FE/HE \cite{bian2021privacy} & \cmark &  \cmark/\cmark & M-H & M-H & Good & H & H  & Same as HE/FE \newline Side-channel risk & TEE: Non-linear operations \newline HE/FE: Linear operations   \\
    & \centering SE/ORAM+TEE \cite{hoang2019hardware} & \xmark &  N/A & N/A & N/A & N/A & L-M & L-M  & No leakage \newline Obliviousness & Partial obliviousness \footnote{}\\
    \bottomrule
  \end{tabularx}
  \begin{threeparttable}

\begin{tablenotes}[flushleft]
    \item[\dag] This table employs the following abbreviations: {\textbf{Comp}}: computation, {\textbf{Comm}}: communication, and {\textbf{L, M, H}}: Low, Mid, and High, respectively, \textbf{Avg}: Average, \textbf{\cmark,\xmark}: Supports and not supports respectively, and \textbf{N/A}: Not applicable. \textbf{\tnote{1}} PET type: {\textbf{C}}: Cryptography-based, {\textbf{H}}: Hardware-based, {\textbf{L}}: Learning paradigm-based, {\textbf{S}}: Statistical-based, and {\textbf{Hybrid}}: Some notable hybrid designs.
    \textbf{\tnote{2}} Floating-point to fixed-point number conversion causes loss of precision and accuracy degradation.
    \textbf{\tnote{3}} In semi-honest security, parties execute the protocol correctly but can be curious to obtain additional information. In malicious security, parties may also deviate from the execution of the protocol. Moreover, MPC can provide $(t, n)$ threshold security where the collaboration of $t$ or more of the parties is required for the computation \cite{shamir1979share}. However, in any case, the risk of colluding servers is present.
    \textbf{\tnote{4}} The main properties of ZKP are Completeness, Soundness, and Zero-knowledge. (See Section \ref{sec:zkp}).
    \textbf{\tnote{5}} While HE and FE can be employed in training scenarios, their primary utilization focuses on the inference stage.
    \textbf{\tnote{6}} This is the main difference with MPC, besides the post-quantum security it provides through approaches like Lattice-based \cite{nejatollahi2019post}, Hash-based \cite{li2022hash}, etc.
    \textbf{\tnote{7}} General SE schemes leak information about files or queries to the server, including details such as the search and access patterns.
    \textbf{\tnote{8}} Obliviousness enables users to access encrypted data on an untrusted server while concealing access patterns. This can have other variants, such as distributed ORAM with threshold collusion resistance presented in \cite{hoang2017s3oram}.
    \textbf{\tnote{9}} Model size reduction can impact accuracy.
    \textbf{\tnote{10}} Setting up and managing the enclave poses overhead.
    \textbf{\tnote{11}} In FL, the training process is iteratively performed across the participating nodes and the aggregator until the model converges.
    \textbf{\tnote{12}} FL nodes engage in a multi-party computation to determine the best label for the inference output.
    \textbf{\tnote{13}} FL aggregator combines the model updates from FL nodes in each iteration, leading to a deviation from the accuracy that could be realistically attained.
    \textbf{\tnote{14}} While DP is not employed individually for inference, its integration in a hybrid design with MPC and FL (See Section \ref{sec:fl}) allows for its incorporation, albeit with an augmented cost and reduced accuracy.    
    \textbf{\tnote{15}} Existence of TEE does not require full obliviousness for the search algorithm as the search and update are done in a secure enclave.
 \end{tablenotes}  
\end{threeparttable}
  \vspace{-5mm}
\end{table*}

\section{Private Search, Access, and Retrieval for Medical Images}\label{sec:encsearch} 
In this section, we present secure access, private search, and information retrieval mechanisms that can potentially be integrated into the initial stages of the Deep Radiomics pipeline, namely data storage and retrieval (See Fig. \ref{fig:ppmlclassifcation}).

\vspace{-3mm}
\subsection{Searchable Encryption to Enable Private DR Services}
SE \cite{song2000practical} enables storing data in encrypted form on the remote storage server for various purposes like data backup or data sharing and later performing keyword(s)-based search over them. An SE scheme typically involves a data owner, authorized data user(s), and an untrusted remote server. The data owner encrypts the data before uploading it to the server. An authorized data user sends a keyword search query and a search token to the server, performing the encrypted search and returning the encrypted results. For example, a hospital can store patient data on untrusted servers in encrypted form and use SE to retrieve specific records (e.g., recent X-ray images) for Deep Radiomics analysis (Fig. \ref{fig:mlpipeline}-A). Additional applications of SE in healthcare include IoT healthcare \cite{seaencPMRSS}, content-based image retrieval using feature vectors \cite{seaencMedImg}, and large-scale encrypted medical images using CNNs \cite{guo2020privacy}.

While SE schemes can be classified based on factors such as search structure (index-based, sequential scan), search functionality (keywords, fuzzy keywords, phrases), etc, SE techniques can be broadly divided into two main categories: Symmetric SE (SSE) and Asymmetric SE (ASE). SSE allows only the entity with the symmetric key to generate searchable ciphertexts and search tokens, while ASE enables anyone with access to the decryptor's public key to encrypt data, allowing subsequent searches to be conducted by the private key owner(s) \cite{behnia2018lattice}. In addition, Dynamic Symmetric Searchable Encryption (DSSE) \cite{kamara2021dynamic,cash2014DSSE} extends SE to allow searching and updating over the encrypted data via an encrypted index, offering high efficiency and security for Deep Radiomics. DSSE has been used in Deep Radiomics for content-based image retrieval from Cloud image repositories \cite{seaencPIC,seaebcFerr,seaencXia}.

DSSE integration to Deep Radiomics encounters several challenges. The fuzzy nature of medical data, with minor spelling and typographical errors, complicates keyword searches \cite{zhang2023verifiable}. Implementing fault-tolerant DSSE schemes \cite{wang2014privacy,fu2016toward,hoang2016practical} is crucial to address these issues, particularly in the medical domain \cite{medicalfuzzykeywordsearch}. Additionally, DSSE schemes often struggle with irrational queries caused by human errors, leading to security risks such as repeated addition or deletion of the same entry or attempts to delete non-existent entries. Developing robust DSSE schemes \cite{dou2024dynamic,xu2022rose} to counter these vulnerabilities is essential. While conventional DSSE enhances search and update efficiency, it may compromise privacy by revealing query information to the server through search and access patterns. Adversaries can exploit these leakages using statistical attacks to recover user data and query information. This issue can be mitigated using extended DSSE schemes \cite{zuo2020DSSEforward}, which provide forward and backward privacy. Forward privacy ensures the server cannot link newly updated files to previous queries, and backward privacy prevents the server from learning about files that were added and then deleted between two identical search queries. These privacy guarantees facilitate the secure sharing of EHRs between hospitals \cite{forbackDSSEMedSharing}.

\vspace{-3mm}
\subsection{Mitigating Leakages of Encrypted Search Services}
In addition to requiring forward and backward privacy, higher security measures are needed to eliminate harmful information leakage and mitigate attacks \cite{kornaropoulos2022leakage}. DSSEs should also be equipped to achieve shareability \cite{wang2023keyword} and post-compromise security \cite{chen2024power}. While such security properties ensure the user's data remains private, there are instances where the remotely stored data is public (e.g., on a public file server). In such cases, only the user must know anything about the retrieved data and the query when issuing a query to this storage. The primitives designed for this purpose fall into the group of oblivious retrieval and access primitives. PIR \cite{pirchor} involves an interaction between two entities: a user and a database. The goal is for the user to retrieve a specific record from the database without revealing any information (such as the item's index) to the untrusted database owner (Fig. \ref{fig:mlpipeline}-A). The integration of PIR in Deep Radiomics has been explored by \cite{mayberry2013pirmap}, who developed a proficient PIR solution for retrieving large X-ray and MRI files within a widely adopted cloud computing paradigm (MapReduce).

PIR approaches can be categorized into Information-theoretic PIR (IT-PIR) and Computational PIR (CPIR) solutions. IT-PIR ensures the server remains oblivious to the user's request by requiring the download of the entire database, thereby guaranteeing privacy. The first IT-PIR \cite{pirchor} was introduced in a multiple-server setting where each server holds a copy of the same database and cannot communicate with the others. Later, Kushilevitz et al. \cite{kushilevitz1997replication} introduced PIR in a single-server setting. Two efficient multi-server PIR protocols include XoR-based \cite{chor1998private} and Secret Sharing-based \cite{goldberg2007improving} methods. However, these approaches are impractical for Deep Radiomics due to substantial communication overhead. In contrast, CPIR relies on cryptographic constructions (e.g., using Fully Homomorphic Encryption), which introduce more computational overhead than communication. While PIR can prevent leakage, it lacks support for information updates (write operations) in the database, necessitating the use of ORAM \cite{goldreich1987towards} used in schemes like Oblivious Data Structure Encryption (ODSE) \cite{ODSE,hoang2017s3oram}. 

ORAM is a probabilistic RAM machine initially designed to execute programs and handle data without revealing information through physical memory accesses. It effectively seals access pattern leakages of SE and enhances privacy \cite{hoang2019hardware} (Fig. \ref{fig:mlpipeline}-A). Among the various ORAM schemes, tree-based versions like Path ORAM \cite{stefanov2018path} and its variants, including Circuit ORAM \cite{CircuitORAM} and Onion ORAM \cite{OnionORAM}, are currently the most efficient ones. Practical ORAM constructions tailored for databases with large block sizes, such as medical image databases, have been shown by \cite{mayberry2015practical}. Some schemes have explored integrating server-side computation in ORAM, shifting computational burdens from the client to the server using computationally intensive HE. Additionally, multi-server ORAM schemes \cite{stefanov2013multi} utilize two non-colluding servers capable of computation. Leveraging Shamir Secret Sharing and Multi-party computation (See \ref{sec:mpc}) to mitigate the complexity associated with HE or other cryptographic primitives has led to achieving constant communication overhead for client-server interactions \cite{hoang2017s3oram,chen2022titanium}. Further advancements to exploit the properties of DSSE, PIR, and write-only ORAM resulted in presenting a series of schemes that achieve high security and efficiency over generic ORAM \cite{hoang2017oblivious}.

{\em Solutions like ORAM and PIR emerge from the information leakages inherent in search and update operations, but their computational costs present a hurdle for Deep Radiomics. Therefore, developing an efficient, user-friendly, and secure searchable scheme tailored for Deep Radiomics is crucial. Additionally, existing schemes predominantly focus on searching over a single data type (text), with only a few supporting multiple data types. However, supporting various data types is advantageous, particularly in Deep Radiomics (e.g., Searching and retrieving patients with a particular property within their MRI scans). Thus, exploring cross-data and cross-index schemes represents a promising direction for future research.}

\section{Executing Deep Radiomics Under Encryption}
In this section, we explore computation under encryption mechanisms that are integrateable into the third and fourth stages of the Deep Radiomics pipeline, which are Deep Radiomics training and inference, respectively (See Fig \ref{fig:ppmlclassifcation}).
\subsubsection{\textbf{{Secure Multi-party Computation (MPC) for Distributed Medical Image Classification (and Decision Making)}}} \label{sec:mpc} 
MPC \cite{yao1982protocols} enables mutually untrusted parties to jointly compute a function on their private inputs, ensuring that no party gains additional information beyond what can be inferred from their input and the function's output. MPC has been utilized for secure distributed medical image analysis and classification \cite{alvarez2020secure,liu2021towards,yu2023pomic}. The first two-party MPC protocol \cite{yao1982protocols} represents the target function as a circuit of logical gates (e.g., AND, OR, and XOR), with one party garbling the circuit and sending it to another party for evaluation (Garbled Circuit (GC)). The circuit evaluation is further completed using Oblivious Transfer (OT) \cite{OTRabin}, ensuring that the holder remains unaware of the requester's selections and the requester acquires no knowledge of the unchosen garbled circuit inputs. Additionally, Secret Sharing (SS) \cite{shamir1979share} can distribute a secret among multiple parties, allowing them to reconstruct it collaboratively. Further, using the shared circuit evaluation approach, parties can evaluate the circuit on the shares of their inputs and obtain shares of the function's output, which can then be reconstructed by a sufficient number of parties \cite{BGW}.

While constant communication rounds characterize GC, the performance may degrade with more AND gates in the circuit representation. Strategies like replacing non-XOR (NXOR) gates with XOR gates \cite{kolesnikov2008improved} can enhance circuit efficiency. However, applying GC to Deep Radiomics may lead to significant computation and communication costs despite theoretically supporting linear and non-linear computations. For example, CNNs with numerous inputs per layer require dedicated comparison circuits for simple activation functions like ReLU \cite{CNNsurvey}. Although the boolean nature of GC is suitable for specific NNs like Binary Neural Networks (BNN) and Ternary Neural Networks (TNN), where model parameters values belong to the set \(\left\{-1, 0, 1\right\}\), this simplification introduces overhead in CNNs with floating-point parameters. Thus, arithmetic circuits are preferred for prevalent CNN operations like addition and multiplication.

In contrast to GC, OT protocols impose significant communication overhead, which scales with the input bit length due to public-key cryptography. To optimize OT for linear computations like matrix multiplication, correlated OT can be employed for the dot product. In this method, the dot product results from aggregating multiple elementwise multiplications, and the shares of the dot product are derived by combining the corresponding shares of elementwise multiplications at two parties. While BNNs exhibit efficient OT protocols as their binary model parameters need an input bit length of one, CNNs pose challenges as parameters are multi-bit length floating-points, resulting in significant communication overhead and potentially compromising model accuracy. Although optimizations allow for over ten million OT executions per second, performance considerations remain essential, like offloading input-independent computations to the offline phase.

SS enables CNN linear computation through a two-phase process, encompassing input-independent offline and input-dependent online computations (Fig. \ref{fig:mlpipeline}-E). In the offline phase, a multiplication triplet is established using OT between a data owner and a cloud server, with each party holding distinct shares of these three values. Subsequently, the dot product is efficiently computed using the precomputed shares in the online phase. While SS proves effective in low-delay settings, it may be less favored in delay-sensitive scenarios due to the computational load in the offline phase.

For non-linear functions such as the ReLU activation function and max-pool function (common in pooling layers) \cite{CNNsurvey}, the efficient computation can be achieved through GC. These implementations avoid approximation and achieve performance comparable to the original Deep Radiomics models. However, activation functions like Sigmoid \cite{CNNsurvey} can be approximated using ReLU variants, introducing an accuracy-efficiency tradeoff for CNNs. While overhead may be reduced through pipelining techniques based on parallel computation, such approximations may still incur computational costs. Moreover, as number comparison involves comparing with zero and evaluating the most significant bit (MSB), SS can be utilized for efficient MSB evaluation, enabling approximation-free computation for functions like ReLU and max-pooling. Although splitting the input of non-linear functions to derive partial non-linear results and combining them to get the output values is feasible, SS-based approaches typically require more communication rounds than GC-based ones (Fig. \ref{fig:mlpipeline}-E).

Many Deep Radiomics model training and inference protocols combine multiple secure computation techniques. SecureNN \cite{wagh2019securenn, wagh2020falcon} and Falcon \cite{wagh2020falcon} exclusively utilize SS for three-party computations. Chameleon \cite{riazi2018chameleon} employs both GC and SS for two-party computations, while Trident \cite{chaudhari2019trident} utilizes GC and SS for four-party training or inference scenarios. Furthermore, integrating with MPC enhances resilience against common ML attacks. For example, Property Inference Attacks (PIA) \cite{ateniese2015hacking}, aimed at extracting implicit properties unrelated to the learning task, are mitigated through a secure two-party training approach involving two non-colluding servers receiving distributed private data from the data owner \cite{mohassel2017secureml}. Additionally, Membership Inference Attacks (MIA) \cite{shokri2017membership}, a black-box attack seeking to determine if an input sample was part of the training set based on model output, can be addressed through MPC-based summation of vectored data during federated learning updates \cite{bonawitz2017practical}.

{\em Hybrid designs show promise for encrypted execution, often outperforming single-primitive designs despite conversion overhead (e.g., arithmetic to boolean shares). Although single-primitive designs avoid this, individual primitive's suitability for specific functionalities (linear or non-linear) is undeniable. Hence, developing efficient single-primitive designs without conversion overhead is crucial. Furthermore, employing batching techniques can enhance data processing rates, which is achievable through pipelining for GC.}

\subsubsection{\textbf{{Homomorphic Encryption (HE) for Encrypted Radiomics (in Single-server setting)}}} \label{sec:he} 
HE enables the encryption of plaintext into ciphertext, allowing certain operations to be performed on the ciphertexts while producing the same results as if applied to the plaintexts (homomorphism) without the need for decryption beforehand. HE supports primary operations like addition and multiplication and is categorized based on the operations it supports: Partially HE \cite{rivest1978data} supports only addition, Somewhat HE (SHE) \cite{boneh2005evaluating} allows for an unbounded number of additions and a single multiplication, Leveled HE (LHE) \cite{fan2012somewhat} permits a predetermined number of additions and multiplications, and Fully HE (FHE) \cite{chillotti2020tfhe} supports an unbounded number of operations.

HE's integration with Deep Radiomics involves encrypting data before transmission to the server for model training or inference, ensuring computation privacy. After performing the desired computation on the encrypted data, the result is communicated in encrypted form, ensuring comprehensive output privacy. Works \cite{yang2023dynamic,zhang2022homomorphic} demonstrated the use of HE in FL for balancing costs and privacy in medical imaging and IoT healthcare. For HE-based inference, BNormCrypt \cite{chabanne2017privacy} and CryptoDL \cite{hesamifard2018privacy} rely solely on LHE, while others such as GAZELLE \cite{juvekar2018gazelle} use GC in addition for more robust privacy.

Most operations in Deep Radiomics, such as weighted sums, matrix multiplication, and dot products, are linear and compatible with HE. However, computing non-linear operations like activation functions (e.g., ReLU, Tanh \cite{CNNsurvey}, and Sigmoid) with HE presents challenges due to its limited addition and multiplication operations, resulting in high computation complexity. Additionally, using HE with CNNs can impose multiplicative depth limitation, which restricts the number of consecutive multiplications before the ciphertext can no longer be decrypted correctly. Furthermore, training CNNs with HE involves forward and backward passes, leading to significant noise accumulation and exacerbating the impact of multiplicative depth, particularly during training. These challenges primarily restrict HE usage to inference rather than training in Deep Radiomics applications (Fig. \ref{fig:mlpipeline}-F).

In CNNs, convolution layers involve matrix multiplications and dot products, with each matrix element corresponding to a HE-compatible ciphertext. However, the standard implementation of $d$-dimensional matrix multiplication requires $d^3$ multiplications, and the dot product of two $d$-dimensional vectors necessitates $d$ multiplications. To enhance multiplication efficiency, strategies like shift operations can be applied, exploiting the property of binary format where multiplying by a power of two is akin to shifting the decimal point. Nonetheless, shift operations mandate quantizing CNN weights to powers of two and representing CNN inputs as fixed-point binary numbers, as demonstrated in CryptoNets \cite{gilad2016cryptonets}. Alternatively, Single Instruction Multiple Data (SIMD) operations, also known as ciphertext packing, can be utilized to reduce ciphertext count, decrease latency, and optimize data organization. While most HE schemes support SIMD operations, TFHE is an exception and cannot be employed in these packing schemes. Moreover, the absence of universal packing schemes has rendered efficiently handling data for operations like convolutions challenging.

Pooling layers, like convolution layers, employ pooling functions across their windows. However, max-pooling, the prevalent choice, proves inefficient with most HE schemes, except for Scale-Invariant HE \cite{lou2019she}. Consequently, many propose to replace max-pooling with (scaled) average-pooling. A prevalent scaling factor is the number of input elements, which converts average pooling into input summations without multiplication, offering tighter control over values' magnitude. While average pooling preserves output within input range values, scaled average pooling may yield larger outputs, necessitating careful HE parameter selection to ensure all values fit into the input space.

Computing non-linear functions like activation functions varies depending on the specific application. For instance, in Scale-Invariant HE, the TFHE library \cite{tfhelib} is employed to implement ReLU, while a step function is implemented using customized bootstrapping operations within TFHE as demonstrated by \cite{bourse2018fast}. However, in FHE, operations on encrypted data typically increase the noise level in the ciphertext, potentially posing a security threat if the noise becomes excessive. To address this, bootstrapping is used to refresh the ciphertext, reduce noise, and restore security. Each activation function evaluation is essentially a bootstrapping operation, facilitating the construction of deep networks with unlimited depth. Nonetheless, training a network with non-standard activation functions like a step function can be challenging due to their lack of gradient information. Consequently, schemes like CryptoRNN \cite{cryptornn} delegate activation function computation to the client, allowing flexibility in choosing any desired function. Moreover, in hybrid designs \cite{juvekar2018gazelle,huang2022cheetah}, HE can be used for linear functions and MPC for computing non-linear activation functions. Moreover, integrating with Intel SGX (Section \ref{sec:tee}) can enable computing HE-unfriendly functions \cite{wang2019toward,coppolino2020vise,xiao2021privacy}.

HE-based Deep Radiomics constructions defend against privacy attacks such as Reconstruction Attacks (RA) \cite{zhang2020secret}, which aim to partially or fully reconstruct one or more training inputs, with or without their corresponding training labels. Attacks discussed in \cite{chang2020attacks} against perceptibly encrypted images in Deep Radiomics models demonstrated that encrypted images retain some information about the original image, albeit invisible to humans. However, the semantic security of HE makes such attacks' success impossible. Similarly, MIAs are ineffective because the same argument applies to RAs.

{\em Current HE schemes face challenges in supporting operations like activation functions (e.g., Sigmoid) and conducting comparisons or selecting maximum/average values within a set, which impacts tasks like average pooling. Solutions include polynomially approximating activation functions, input normalization, and pre-computing functions using homomorphic table lookups. However, the computational costs of HE increase with data input and CNN depth, hindering its practicality for Deep Radiomics. Strategies like employing low-degree polynomial approximations and reducing CNN layers are considerable but may compromise accuracy, disfavoring precision medicine. Moreover, batching and utilizing hardware accelerators (e.g., GPUs) offer potential improvements, though memory constraints require novel encoding and packing schemes. Hence, hardware-software co-design to enhance encrypted data pipelining is a vital research direction.}
\subsubsection{\textbf{{Functional Encryption (FE) for Selective Medical Image Diagnosing}}} \label{sec:fe}  
In a conventional public-key encryption scheme, the recipient decrypts data fully using a private key. However, controlled or restricted access to plaintext is sometimes necessary, allowing only specific individuals to decrypt the ciphertext. For instance, a hospital might encrypt patient data and send it to a third party for heart disease trend analysis. To protect patient privacy, a mechanism is needed that allows the third party to selectively access only certain parts of the encrypted data, such as recent prescriptions.

FE \cite{feformalized} uses specialized secret keys to decrypt the output of specific functions applied to encrypted data without revealing the actual inputs. The application of FE in Deep Radiomics for medical primary diagnosis has been explored by \cite{hua2020camps, cui2023medical}. FE-based Deep Radiomics employs two main approaches: Inner-product FE (IPFE) and Quadratic FE (QFE). IPFE supports both training (including forward and backward propagation) and inference by calculating the inner product of an encrypted vector \(\textbf{x}\) and a plaintext vector \(\textbf{y}\) without decrypting \(\textbf{x}\). For activation functions, it computes the inner products of encrypted inputs and weight matrices. QFE, specialized for faster inference, uses polynomial approximation on encrypted data within polynomial neural networks, suitable for linear components like fully connected layers, convolutions, average pooling, and polynomially approximated activation functions.

Despite its use in PPDR (Fig. \ref{fig:mlpipeline}-F), FE faces significant limitations and challenges. Current FE approaches, including linear IPFE and quadratic QFE, are constrained in their capabilities. For example, IPFE-based CryptoNN supports training and inference for a five-layer neural network, while QFE-based protocols enable inferences for a two-layer network. IPFE methods, suited for simple computations like inner products, struggle with the complex requirements of CNNs in medical imaging and require numerous ciphertexts for convolutions. QFE approaches are restricted to polynomial NNs of up to five degrees, which is insufficient for complex CNN architectures like VGGNet and GoogleNet. FE constructions also depend on computationally costly cryptography, leading to performance overheads that, while reasonable for partially encrypted tasks, are inadequate for complex learning tasks over large datasets. 

{\em The limited functionalities of current FE schemes and the computational and memory costs of ciphertexts and keys make implementing complex CNNs for Deep Radiomics challenging, especially for large datasets like ImageNet. Additionally, no FE scheme leverages hardware accelerators (e.g., GPUs), vital for CNNs. These shortcomings highlight the motivation for efficient FE schemes supporting GPU acceleration, particularly for the training phase in Deep Radiomics.}

\section{Verifiable Execution of Deep Radiomics}\label{sec:zkp}
ZKP\cite{goldwasser2019knowledge} is an interactive system where a computationally unbounded prover attempts to convince a probabilistic polynomial-time verifier of a statement's truthfulness. ZKP systems must satisfy three critical properties: (1) Completeness, ensuring the verifier always accepts the proofs of honest provers for true claims; (2) Soundness, ensuring rejection if a malicious prover presents proof for false claims; and (3) Zero-Knowledge, guaranteeing confidentiality during the protocol.

The challenges posed by interactive protocols, especially when the verifier is not consistently online or when communication bandwidth is limited, highlight the need for non-interactive ZKPs (NIZKPs) \cite{blum2019non}. In NIZKPs, the prover generates the proof once and sends it to the verifier in a single round, overcoming these limitations. Additionally, the desire for concise proofs further drives the adoption of Succinct Non-Interactive Arguments of Knowledge (zkSNARKs) \cite{bitansky2012extractable}, which are highly preferred over traditional ZKPs for Deep Radiomics due to their efficiency and effectiveness.

ZKPs play a vital role in PPDR by verifying model ownership and inference. For instance, in \cite{zkpintelproperty}, ZKPs were used to verify ownership of a Deep Radiomics model designed for removing bones from chest X-rays. Integrating ZKPs such as zkSNARKs into Deep Radiomics involves a verifier with limited computational capabilities delegating training or inference responsibilities to a more capable entity with mutual agreement on tasks, datasets, and models. Combining ZKPs with commitment schemes \cite{pedersen1991non} enhances model security and establishes non-repudiation by safeguarding the confidentiality of Deep Radiomics computations. The server (prover) commits to learning elements, executes the task, and produces a proof. After completion, the prover transmits the result, commitment, and proof to the verifier for validation check (Fig. \ref{fig:mlpipeline}-H).

ZKPs defend against attacks like data poisoning \cite{liu2018trojaning}, where attackers manipulate training data (e.g., poisoning features, changing model weights, etc.) to influence learning models. While prior research focused on preventing such attacks or concealing model updates, Nguyen et al. \cite{nguyen2023preserving} presented a balanced privacy and detection scheme using ZKPs. Defending against MIAs using the ZKP-based framework was presented by \cite{tang2023pile}, which protected local gradients and global models' privacy through gradient verification.

{\em Deep Radiomics models relying on floating-point numbers pose challenges for using ZKPs, as they operate over integers. This incongruence leads to significant computational overhead in proof generation. For example, using Groth16 \cite{groth2016size} for VGG16 requires ten years of computation and 1000 terabytes of storage. Suggested solutions include tailoring proof systems for specific computations, utilizing hardware accelerators and pipeline designs to expedite proof processes, and balancing security and efficiency through multi-round training in which a random subset of rounds are verified. Building efficient ZKP protocols tailored for Deep Radiomics is an open problem.}

\section{Collaborative Learning Paradigm for Deep Radiomics and Hybrid Designs}\label{sec:fl}
Traditional learning methods often involve transferring data and models to third parties when computational resources are inadequate. Moreover, data providers (e.g., hospitals) may lack sufficient training data, resulting in models with limited accuracy. To overcome this, utilizing Generative Adversarial Networks (GANs) \cite{che2017boosting} for data augmentation is suggested. Yet, GANs may not consistently provide the diversity needed for healthcare systems. Another approach is exchanging data between providers for training, but the large size of Deep Radiomics images can lead to communication overhead and scalability issues. Additionally, diverse privacy policies among providers make this approach impractical.

FL \cite{adnan2022federated} is a paradigm for collaboratively training a global model across multiple datasets distributed over separate nodes without explicit data sample exchange. In FL, each node sends its locally trained model parameters to a central node for aggregation into the target model. The globally aggregated parameters are then redistributed to the nodes, and this iterative process continues over multiple rounds until the desired accuracy level is reached (Fig. \ref{fig:mlpipeline}-D). While FL offers benefits like generalizability and scalability, it often involves a trade-off, resulting in a loss of model accuracy.

FL is widely applied in medical imaging, with collaborative endeavors expanding \cite{ng2021federated}. For instance, in federated brain imaging for tumor segmentation based on functional MRI (fMRI) \cite{li2020multi}, deep NNs (DNNs) are locally trained on datasets integrated with Differential Privacy (DP) \cite{wu2021incentivizing,li2019privacy} to mitigate reconstruction attacks through artificial noise addition and further aggregated. Notably, examples of such integration include \cite{adnan2022federated} for histopathology image analysis (e.g., lung cancer) and \cite{li2019privacy} for brain tumor segmentation. Moreover, FL assists in pathology images \cite{zhang2022splitavg}, automating tissue region segmentation and embedding into low-dimensional features using CNNs \cite{lu2022federated}.

Despite being capable of PPDR, FL faces numerous challenges. Transmitting locally trained parameters to a central node poses privacy risks, as updates may contain sensitive information susceptible to reconstruction attacks. One solution involves integrating DP into FL by adding noise to the local dataset \cite{wu2021incentivizing}. FL is also vulnerable to MIAs and poisoning attacks, where malicious or low-quality data can degrade accuracy or create trojan-enabled models. Additionally, if datasets are of low quality or contain various types of medical data, like images, audio, and text, the accuracy of the final model may suffer. These challenges can be addressed through heterogeneous FL approaches, private ensemble learning methods capable of handling diverse data types, and reputation-based mechanisms ensuring nodes contribute high-quality, trustworthy data and updates to the training process.

While capable of standalone operation, FL can benefit from hybrid designs to effectively address challenges as well. Integration of HE into FL for various medical imaging tasks has been explored, including lesion classification \cite{yang2023dynamic}, cancer image analysis \cite{truhn2024encrypted}, and U-shaped medical image networks \cite{yang2023dynamic}. Furthermore, FL and MPC have been applied in healthcare for tasks like diagnosing pneumonia \cite{siddique2023privacy}, analyzing histopathology images \cite{MPCFL}, multi-institutional medical imaging \cite{kaissis2021end}, and multi-party FL inference systems for Diabetes Mellitus risk prediction \cite{su2023multi}. Moreover, using secure voting via MPC, nodes can perform secure voting by adding DP noise to their local inferences and collectively decide on the final inference label \cite{kalapaaking2022smpc} or aggregated model parameters.

{\em FL's privacy challenges, such as MIAs, unintentional information leakage, and generative adversarial networks due to the high sensitivity of health-related data, are undeniable. Practical solutions involve hybrid and scalable FL designs with other PETs like DP. Advanced cryptographic methods such as MPC and incentive mechanisms based on Game theory \cite{flgametheory,flgamethoerytwo} and Blockchain \cite{flBlockchain,flBlockchaintwo} offer promising directions for protection.}

\section{Trusted Hardware for Efficient Deep Radiomics and Hybrid Constructions}\label{sec:tee}
Using cryptographic primitives for confidential computing introduces computational and communication overheads. A more efficient approach involves leveraging hardware-based TEEs, which operate independently of the operating system or hypervisor. TEEs facilitate isolated and verifiable code execution within secure enclaves (protected memory regions) and differ from Trusted Platform Module (TPM), where secure hardware is physically isolated from other modules. The benefits of allowing smooth transitions between normal and trusted modes of operation have led commercial processor companies to integrate TEEs into their latest processor designs. Notable TEEs include Intel Software Guard eXtensions (SGX) \cite{intelsgx} for processor-based TEEs, virtual-machine (VM)-based TEEs such as AMD Secure Encrypted Virtualization (SEV) \cite{amdsev}, ARM TrustZone \cite{armtrustzone}, and RISCV Keystone \cite{riscvkeystone}. Moreover, efforts are underway to integrate TEEs with ML hardware accelerators (e.g., GPUs, TPUs \cite{tpu}, and NPUs \cite{npu}), which is critical, especially for Deep Radiomics due to the increasing complexity of NN training and the expansion of datasets. Notable integrations of TEEs and accelerators include Graviton \cite{graviton} (with GPU), Vaswani et al. \cite{graphcoretee} (with Graphcore), and the NVIDIA H100 Tensor Core GPU \cite{nvidiaNVIDIAH100}.

Establishing trust between TEEs and external computing nodes is vital for ensuring privacy. Root of Trust (RoT) measurement verifies a TEE's integrity by examining critical system components before executing stored code. Remote Attestation (RA) validates the TEE's integrity when a computing node transmits code. Through attestation, the node confirms the remote TEE's trusted state and secure execution of the intended code. This process includes signing a RoT measurement report, verifying the signature, comparing it against a trusted reference, and generating an attestation certificate, which is then transmitted to the attester.

To utilize TEEs for PPDR, data and model owners must transmit their assets to a remote TEE situated on an untrusted computing node (Fig. \ref{fig:mlpipeline}-B). When a node serves as a server providing training or inference services, the data sent to this node must be stored within the TEE to ensure privacy. Similarly, if another entity owns the Deep Radiomics model, it should also be provisioned to the TEE. However, for inference purposes, it is acceptable for the model not to reside in the enclave if the server is untrusted or not the model owner, as inference does not impact the model parameters. Conversely, when the computing node is a client with private data stored locally, another entity may own the model as intellectual property. Examples of such inference scenarios include on-device inference, on-device personalization, and FL, where the model is trained locally and aggregated with other locally trained models \cite{mondal2021poster,mo2021ppfl}.

Despite TEEs offering superior performance over cryptography-based solutions for PPDR, several challenges persist. Resource constraints hinder deploying the entire learning pipeline, covering data preparation and storage, training, and inference within TEEs. For instance, adversarial attacks, such as adding noises to the inputs and their labels either by physical surroundings or by attackers (side-channel attacks), pose challenges for TEEs to control the first step of the Deep Radiomics pipeline, which is data generation and preparation. Moreover, assigning training and inference tasks to TEEs on an untrusted host does not address privacy breaches, as transmitting final results requires potentially secure channels. While safeguarding model parameters in TEEs can counter reconstruction attacks, it may be ineffective against attacks like MIAs, where information leakage occurs from the model's output. 

Considering TEEs' benefits and challenges, a common hybrid strategy integrates TEEs with other solutions for ease of deployment and performance advantages. For instance, Intel SGX can offer secure encrypted searches while protecting against side-channel attacks and access pattern leakages. Hardware-assisted ORAM, utilizing SGX, can reduce bandwidth overhead and allow outsourcing of large databases \cite{hoang2019hardware,hoang2020mose}. Integrating TEEs with SSE (e.g., for image-based apps) can eliminate the need for complete obliviousness and ensure queries don't reveal document connections or expose data updates \cite{shaon2020sgx,amjad2019forward}. Additionally, Intel SGX can be combined with HE in medical image segmentation for brain and heart disease analysis \cite{bian2021privacy}, and with MPC to allow for secure parallel key exchanges \cite{bahmani2017secure,choi2019hybrid,felsen2019secure}.

{\em A promising research direction is in-memory computing (e.g., matrix multiplication), which eliminates data transmission between CPUs, GPUs, and memory. This capability can achieve significant speedups, up to 100$\times$, and can be exploited, especially with recent TEEs with large memory sizes. Moreover, while a single TEE may not suffice to safeguard the entire Deep Radiomics pipeline, employing multiple TEEs can offer comprehensive protection. Integrating a multi-TEE design with verification mechanisms such as Blockchains or third-party proxy attestation or using it with other PETs like MPC presents a promising approach for enhancing the security of Deep Radiomics.}

\section{Statistical Perturbation for Adaptable Tuning of Privacy}
DP \cite{dwork2006differential} provides a systematic framework for assessing the privacy guarantees of a protocol. Initially designed to anonymize the results of interactive database queries, DP prevents an adversary from determining whether a particular data record is in the database by introducing noise perturbation to the queries or the database itself. This noise transforms the analysis outcome into an estimate rather than a precise computation on the original dataset \cite{wang2023differential}. The unpredictability of the added noise results in varying outcomes across different DP analysis runs. A significant challenge, however, in applying DP to Deep Radiomics is balancing privacy and accuracy while preventing accuracy reduction.

DP can be implemented in two main ways: Central DP (CDP) and Local DP (LDP). In CDP, the data curator, who holds the database, is trusted, and random noise is added after collecting data from users. In contrast, in LDP, where the data curator is not trusted, each user perturbs their data before sending it to the curator. Applying DP in Deep Radiomics protects training data against attacks like MIAs when releasing model parameters. Notable integrations are those mentioned in \cite{lu2022federated} for gigapixel whole slide images in pathology, \cite{li2020multi} for multi-site fMRI analysis, and \cite{sheller2019multi} for brain tumor segmentation.

In centralized learning models, applying DP can occur at different levels: gradient (protecting successive model updates), input (training dataset or target function), or label (protecting the learned model). The choice depends on the optimization of the target function. For convex target functions, the noise magnitude of DP is determined by the sensitivity of the learning algorithm. However, due to multiple local minima, sensitivity analysis is impractical for non-convex objective functions, typical in Deep Radiomics. Therefore, noise is added to the gradients instead (Fig. \ref{fig:mlpipeline}-G).

In distributed learning models like FL, privacy is partially preserved as users don't share their private data with the model aggregator. However, sending updated parameters still risks data leakage. DP can mitigate this in three ways: (1) Users apply noise to their updates in the LDP model or integrate DP with SGD. Note that LDP may not ensure user-level privacy with highly correlated data points, (2) the CDP model hides individual clients by trusting the central manager, who perturbs model updates, (3) users train local models independently, add noise to labels, and participate in MPC-based voting for the final label (Fig. \ref{fig:mlpipeline}-C).

Applying DP to Deep Radiomics serves as a defense against various attacks. Traditional methods to counter MIAs often involve retraining models, which is time-consuming. However, leveraging DP with SGD demonstrated by \cite{dpattack,shokri2017membership} significantly reduces the success rate of MIAs. Moreover, studies like \cite{song2013stochastic,abadi2016deep,mcmahan2017learning} have shown that DP with SGD can effectively guard against Model Extraction Attacks (MEA) \cite{tramer2016stealing}, which aim to extract information for complete model reconstruction.

{\em While DP can be applied to input data, model parameters, and model outputs, existing schemes primarily utilize DP with gradients which can accumulate and lead to excessive noise levels, especially in models with many parameters. To mitigate this, objective perturbation has emerged as a promising technique for PPML. However, applying this method to Deep Radiomics encounters challenges due to the non-convex nature of CNNs and the lack of closed-form expressions for their objective functions. Although convex approximation of the loss function is a potential solution, effective strategies to address this issue are still under active investigation.}

\section{Conclusion}
Given the power of automated data analysis through ML, the NIH has identified Radiomics as an effective method for analyzing medical images accurately and efficiently to meet precision medicine requirements. Deep Radiomics leveraging DL has yielded even more effective results. Given the sensitivity of medical images, using PETs to ensure privacy throughout the Deep Radiomics pipeline—from data generation and collection to model training, inference, and release—is essential, especially when outsourced. Therefore, specialized studies are needed to investigate and systematize the effective integration of PETs into the Deep Radiomics pipeline. In this paper, we classified existing PETs and presented practical hybrid PET constructions along with a taxonomy illustrating their potential integration potentials and a comparative analysis detailing assumptions, architectural suitability, and security. Additionally, we offered comprehensive yet broadly accessible technical insights into the potential challenges and strategies for combining PETs at different stages of the Deep Radiomics pipeline. Lastly, we proposed future research directions, identifying challenges and suggesting solutions to enhance the practicality of PETs in Deep Radiomics.

\bibliography{ref}
\bibliographystyle{IEEEtran}
\vfill

\end{document}